\documentclass[a4paper]{article}
\begin{document}

\title{Democritus as Taoist}
\author{Alejandro Rivero\thanks{Zaragoza University at Teruel.  
           {\tt arivero@unizar.es}}}

\maketitle

\begin{abstract}


They identify the full or solid with "what is," and the void or rare with
"what is not" (hence they hold that "what is not" is no less real than "what is")
[Arist. Metaphys. A 4 985b4]

\end{abstract}

J. Needham joined explicitly with Berthelot in remarking the atomism
does not appear anywhere in the alchemical treatises, neither in the
East nor in the West. Which was puzzling, especially because western
alchemists claimed to follow Democritus. In general it was noticed that
other Western philosophical schools can be mirrored in the East, but
atomism seems to be absent.

Perhaps it is time to note that atomism, i.e. the theory of atoms and
molecules in finite size and multiple integer proportions\footnote{We
should call  "neoatomism" to the one culminating in
John Dalton, but the damage is already done}, does not appear
in Democritus himself. Thus it is not so surprising if the followers did not use
it after all.

On the contrary, the people who labels themselves as "Democritians" relies
heavily in the use of principles of duality, as the one quoted above in
the abstract. The ancients thought that atomism was a fusioned theory,
were Leucippus was a disciple of Zeno while Democritus was disciple of
the (mid-)eastern Ostanes. They probably felt this conclusion was supported in
the fact that theological dualism was also an eastern product, from Zoroaster.

The dual scheme of Democritus appears more clearly in the greek
wording. For the first phrase in the abstract above we can transliterate
{\bf pl\^eres kai stereon to on, to de kenon to m\^e on}.
In this language "void" is expressed with the word {\bf kenon}, in a cunning
play to stress the equivalence with "not being", {\bf m\^e on}. Aristotle repeats
this sentence when doing a fast review in [Physics I 5, 188a], so we are pretty
sure it is a basic piece of atomism. As sure the Hellenistic alchemists were;
they understood the duality of cinnabar as other instance of the same kind
of reasonment. Meanwhile, in the East, alchemists had a much better
explicit option: Yin/Yang\footnote{The TTC itself, being an early book, uses
Ying and Yang only once, chapter 42}.

Was there a window of opportunity for mutual influence? The answer seems
affirmative to me.
Since the conquest of Bactria by the disciple of Aristotle, Alexander, there was
open a direct line of contact between Chinese and Greek philosophy. Before this,
the Persian empire interposed, the way was only an indirect one, but here we
can rely again on the suggestions of Persian influences on Democritus, in
the age of Xerxes.

From a broad chronology, the composition of the Tao Te Ching (DaoDe Jing) is contemporary to
the development of atomic theory. According the "Warring States Project"\cite{wsp}, the
 chapters of the TTC we have interested on are to be
considered in the most primitive core, Ch 14-15 being approx 340. Of course Guodian
text gives us an upper limit circa 300 bce. On the contrary side,
David H. Li\footnote{who informed us that  he refers to Democritus in the introduction to
a bilingual DDJ he translated recently}
dates the TTC almost one century before Democritus. Someones even are eager to
accept the date of Lao Zi quitting westwards, 484bce. This is right in the centre of target:
Xerxes became king in 486, and atacked Greece in 480\footnote{One wonders if the 484
date for Lao Zi could be indeed based on the Persian events}.

Continuing with [Metaphys A 4 985], while writing \cite{rivero} I asked a friend, versed
on presocratic
greeks, to try to get the "holographic" text that Aristotle is suspected to be copying into
these paragraphs. With great effort, my friend followed the analysis of
the three properties distinctive of atoms and she concluded that Aristotle was weaving
a (linguistic or logic) map from some unmapped description, according the
following table:

\begin{center}\begin{tabular}{lcl}
(WITHOUT MAP)	& & (WITH MAP)\cr
Where they look at, or turn towards &= &Arist. point of view, placement, position, situation \cr
How they sound or how they are measured &=& Arist. how they rest, shape, role\cr
Whom they touch to, or contact with &=&Arist. relation in a ordered series, order, place, site\cr
\end{tabular}\end{center}

Then, after some time, we found the unmapped terrain accurately enunciated in
the TTC, chapter 14:
\begin{quotation}
"We look for it, but we do not see it: we name it the Equable.

 We listen for it, but we do not hear it: we name it the Rarefied.

 We feel for it, but we do not get hold of it: we name it the Subtle

These three we cannot examine. Thus they are One, indistinguishable".
\end{quotation}

That the body or substrate is One and Same, it is told to us by Aristotle in a
posterior remark, [Metaphys VIII 2, 1042b]:
"Democritus apparently assumes three differences in substance; for he says
 that the underlying body is one and the same in material, but differs in figure,
 i.e. shape; or inclination, i.e. position; or intercontact, i.e. arrangement"
 \footnote{{\bf to men gar hupokeimenon s\^oma, t\^en hul\^en, hen kai tauton,
 diapherein de \^e rhusm\^oi, ho esti sch\^ema, \^e trop\^ei, ho esti thesis, \^e diathig\^ei,
  ho esti taxis}. Online, see {\tt perseus.org} or directly  \cite{rivero}  for the greek texts}.

 To my mind, it is very subtle to decide if the order in the greek  is "listen, look, feel"
or "look, listen, feel". It should be said that my friend affirms she had never seen this
quote, nor any other piece of the Lao Tze book. But even if there was an inconsciently
forgotten influence, it is remarkable.

Now on this light, one could immerse in modern taoist writings and find out pieces of some
atomic flavour (e.g. from the USENET):
\begin{quotation}
The space between things
waxes and wanes, depending upon the things.
Does the space itself actually change?
Does emptiness really exist, on its own? "two different names for one and the same..."
\end{quotation}
but we will refrain here.
This is no news at all, if we notice that even the attribute of indivisibility is sometimes
used when discussing the Tao.  Huang Yuan-Chi (XIXth century) has been translated as
saying "Emptiness and the Tao are indivisible (...) but formless emptiness is of no
use to those who cultivate the Tao". This sentence is not in the line of the theory
of atoms, but it is close. Even the word infinitesimal comes easily in this
state of mind, so lets turn face towards it and some other relationships
coming from mathematics.

Here we should notice in advance that
modern scholars have given no use to the new evidence about Democritus brought
alive by Heiberg  from the Archimedes Palimpsest: Democritus did a non
rigorous calculus for the volume of cone and pyramid. This is surely the
work which Crysippos objects to, in a preserved quote of the stoics.

Lets conjecture that the non rigorous mathematical entities that Democritus
used for this calculus were his {\it a/tomes}. It is not a far fetched conjecture;
before us Cavalieri, without knowing Archimedes attribution, did it... and
he called to his non-rigorous entities {\it in/dividua}.

We know that the problem of the volume of the pyramid was solved in China
by members of the taoist school. The analysis in \cite{dkwagner} shows that the
proof of  Liu Hui uses a taoist vocabulary of the technical points about
evanescent quantities. For instance, Liu Hui argues:

\begin{quotation}
 "The smaller they are halved, the finer [xi] are the remaining [dimensions]. The extreme
 of fineness is called {\it subtle} [wei]. That which is subtle is without form [xing]. When it is
explained in this way, why concern oneself with the remainder?
[Jiuzhang, 168]
\end{quotation}

 And the Heshang Gong, Chap. 14, underscores
"That which is without form [xing] is called {\it subtle} [wei]", and then it elaborates
his commentary on TTC 14 using a very similar vocabulary.

A clue of the use of atoms in the context of a calculation of area can be
got from TTC 11:
\begin{quotation}
"Thirty spokes will converge
In the hub of a wheel;

But the use of the cart
Will depend on the part

Of the hub that is void"
\end{quotation}

Here Lao Tse uses explicitly the expression "not-being" for the void space between
radius!. Be can safely assume that these radius correspond
to "being". A calculation of the weight of a wheel could be done by
understanding that  radius are always in relationship with the
area between. In Democritus, the "being" holds some properties (rhismos,
diathige, trope), and we could conjecture the "non-being" holds others, perhaps
spatial separation or distances.
The whole setup should be very close to modern duality between
differential forms and vector fields.

In geometry the atomists surely felt themselves authorised to keep this
duality all the way to infinite atoms/infinitesimal areas.
 But in physical
evolution, with time as an independent variable, Zeno paradoxes refrained
them... A digression here:
Newton bypassed this issue by mapping Time to a preserved
area (prop. 1 in the Principia) and then taking its limit to zero. Amazingly
this mathematically correct procedure showed to be, three centuries later,
physically false. It is still an open question if the existence of
Plank constant and four elementary fermions relates to some logical
obstruction in the build up of the limit.

And this is all until now.
For the moment we have not gathered more clues nor solid evidence.
Anecdotically, I wonder if it could be possible to locate actual crossed
quotes, assuming some kind of name translation. But,
if Leucippus and Democritus were to be quoted in Chinese, which names should we
expect? If Lao Dan were quoted in Greek, which letters should we expect?
Worse, should the translation cross via a Persian script or dialect? One is afraid
that some regrettable coincidences in initials could happen, adding to confusion:
even Aristotle
shortens sometimes to "L. \& D."  when referring to the atomist forefathers.

As for other possible lines of parallelism,
Chinese versions of Zeno paradoxes are named from time to time\footnote{Zhou (Miller Jew) was kind to
forward me towards \cite[pg. 72 and 32-33]{creativity}, where the arrow paradox is
discussed -but
 context remains undetermined-, as well as some variants of Achilles.
In general I must thank the people
of  {\tt alt.philosophy.taoism} for some pointers during a discussion Jan-Feb 2002, esp Zhou
and Kamerm}
but its
 relationship with taoism is uncertain, to me.
Also, bosonic fields appear in Democritus, {\bf eidola}, but I have not heard of them
in Chinese texts (They are a requisite for interaction if you want to keep always a
vacuum between atoms).

Finally, the reader could object that Taoism has been read into a lot of religions
and philosophical systems, due both to underlying ideas and to simple technical
coincidences, because the word Tao can also be translated as greek "logos"..
Against this we can argue the chronological parallels, the parallels in the posterior
development (alchemy and mathematics) and the textual parallelism.

\end{document}